\documentclass[aps,prc,twocolumn,superscriptaddress,showpacs]{revtex4}

\usepackage{graphicx}% Include figure files
\usepackage{amsmath}
\usepackage{amssymb}
\usepackage{dcolumn}
\usepackage{txfonts}

\begin{document}

% \preprint{}

 \title{
Signature of smooth transition from diabatic to adiabatic states \\
in heavy-ion fusion reactions at deep subbarrier energies}

\author{Takatoshi Ichikawa}%
\affiliation{Yukawa Institute for Theoretical Physics, Kyoto University, Kyoto 606-8502, Japan}
\author{Kouichi Hagino}
\affiliation{Department of Physics, Tohoku University, Sendai 980-8578, Japan}
\author{Akira Iwamoto}
\affiliation{Japan Atomic Energy Agency, Tokai-mura, Naka-gun, Ibaraki
319-1195, Japan}

\date{\today}

\begin{abstract}
 We propose a novel extension of the standard coupled-channels framework 
for heavy-ion reactions 
in order to analyze fusion reactions at deep subbarrier incident energies. 
This extension simulates a smooth transition between
the diabatic two-body and the adiabatic one-body states. 
To this end, 
we damp gradually the off-diagonal part of
the coupling potential, for which the position of the 
onset of the damping varies for each eigen channel. 
We show that this model accounts well for the steep falloff of the
fusion cross sections for the $^{16}$O+$^{208}$Pb, $^{64}$Ni+$^{64}$Ni, and 
  $^{58}$Ni+$^{58}$Ni reactions.
\end{abstract}


\maketitle

Heavy-ion fusion reactions at low incident energies provide a good
opportunity to study fundamental features of the tunneling phenomena in
many-particle systems.
A potential barrier, called the Coulomb barrier, is formed because of a
strong cancellation between the repulsive Coulomb interaction and an
attractive nuclear interaction.
In particular, the potential tunneling at incident energies below
the Coulomb barrier is referred to as the subbarrier fusion reaction. 
One prominent feature of the subbarrier fusion reactions is the large
enhancement of fusion cross sections, as compared to a prediction of 
the simple potential tunneling~\cite{DHRS98}. 
This enhancement has been attributed to the coupling
of the relative motion between the colliding nuclei 
to several intrinsic degrees of freedom, such as a 
collective vibration of the target and/or projectile nuclei.
The coupled-channels (CC) approach, based on this picture, has been 
successful in accounting for the subbarrier enhancement~\cite{Bal98}.  

Because of a recent progress in experimental techniques, 
it has been possible to measure fusion cross sections down to
deep-subbarrier incident energies~\cite{Jiang02,jiang04-2,das07,stef08}. These data 
show a substantial reduction of fusion cross sections at deep-subbarrier
energies from the prediction of the CC calculations that reproduce the
experimental data at energies around the Coulomb barrier, and 
have brought about a renewed interest in this field. 
This phenomenon, often referred to as the fusion hindrance, 
shows a threshold behavior, where the data deviate largely from the
standard CC calculations at incident energies below a certain threshold energy,
$E_s$. 

A key element to understand the fusion hindrance is the
density overlap of the colliding nuclei in the potential tunneling process.
When the incident energy is below the potential energy 
at the touching point of the colliding nuclei, $V_{\rm Touch}$, 
the inner turning point of the
potential is located inside the touching point, and the projectile
is still in the classically forbidden region when the 
two nuclei touch with
each other (see Fig.1 in Ref.\cite{ich07-2}). 
In this situation, the colliding nuclei have to penetrate
through a residual barrier with an overlapping configuration before 
fusion takes place. 
In our previous work~\cite{ich07-2}, we evaluated $V_{\rm Touch}$
using several kinds of ion-ion potential, 
and systematically compared those with
experimentally determined threshold energy $E_{s}$ for many systems. 
The obtained systematics shows
a strong correlation between $V_{\rm Touch}$ and $E_s$, 
indicating strongly 
that the nuclear interaction 
in the overlapping region 
plays a decisive role in the
deep-subbarrier hindrance.

Three different mechanisms have been
proposed so far in order to account for the fusion hindrance.   
Based on the sudden picture, Mi\c{s}cu and Esbensen have 
investigated the
effect of the nuclear interaction in the 
overlap region, in terms of a 
repulsive core due to the Pauli exclusion
principle~\cite{mis06,mis07,Esb07}. Assuming the frozen-density
in the overlapping region, 
they obtained a much shallower potential pocket than the
standard one, which hinders the fusion probability for 
high partial waves. 
Their shallow potential reproduces well the
fusion hindrance. In contrast, we have 
proposed the adiabatic approach by
assuming neck formations between the colliding nuclei in the overlap
region~\cite{ich07-1}. In our model, 
the fusion hindrance originates from the tunneling
of much thicker potential barrier characterized by the adiabatic
one-body potential. 
This model achieved comparable good
reproduction of the experimental data to the sudden model. 
The third mechanism, suggested recently by Dasgupta {\it et
al.}, is the quantum decoherence of channel wave functions caused by the
coupling to the thermal bath~\cite{das07}.
A model calculation based on this picture shows a possibility of the
gradual occurrence of hindrance in 
subbarrier fusion reactions~\cite{dia08}.

Among those three mechanisms, the origin for the deep-subbarrier hindrance
is considerably different from each other. 
The recent precise data for the $^{16}$O+$^{208}$Pb fusion~\cite{das07} 
may provide an adequate system 
to discriminate among the various models, 
because the behavior of its astrophysical S-factor is difficult to
reproduce within a simple model calculation. 
In the model of Esbensen and Mi\c{s}cu, not only 
the collective inelastic channels but also 
the particle transfer channel with modified
coupling strengths are necessary for 
a fit to the experimental data~\cite{Esb07}.
In the estimation of Dasgupta {\it et al.}, it was impossible to obtain
an overall fit to the experimental data from the above-barrier to
deep-subbarrier regions
with a single parameter set for 
the nuclear potential~\cite{das07}.
On the other hand, the performance of the adiabatic model has not yet been
studied for this system, although the concept of the adiabatic potential 
was proved helpful in
the analysis of the potential inversion method in the deep-subbarrier
fusion~\cite{Hagi07}. 

In this paper, 
we attempt to study the deep-subbarrier fusion for the
$^{16}$O+$^{208}$Pb reaction based on the adiabatic model.
Our previous model has a defect that the full quantum treatment for the
two-body part suddenly switches to the semi-classical one for the
adiabatic one-body part, which introduces arbitrariness for the choice
of the Hamiltonian.
To avoid the shortcoming, we develop below
a full quantum mechanics where the CC approach in the two-body
system are smoothly jointed to the adiabatic potential tunneling 
for the one-body system, resulting in an overall good agreement for
the $^{16}$O+$^{208}$Pb reaction, as well as for the $^{58}$Ni+$^{58}$Ni
and $^{64}$Ni+$^{64}$Ni systems.

\begin{figure}
\includegraphics[keepaspectratio,width=\linewidth]{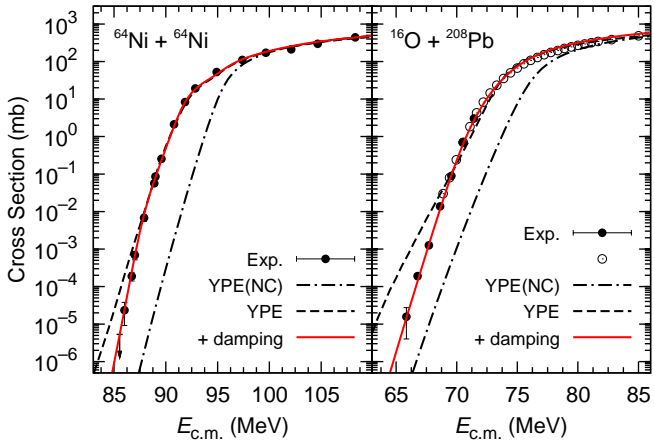}\\%
\caption{\label{fig2} (Color online) Fusion cross sections for
 the $^{64}$Ni+$^{64}$Ni and $^{16}$O+$^{208}$Pb systems. The solid and
 the dashed lines are the calculated result with and without the damping
 factor, respectively.
 The dot-dashed line is the result of no coupling with the YPE potential.
}
\end{figure}

We employ the incoming wave boundary condition in order to simulate 
a compound nucleus formation. 
In order to construct an adiabatic potential model 
with it, 
we postulate the followings: (1) Before
the target and projectile nuclei 
touch with each other, the standard CC model in
the two-body system works well. (2) After the target and projectile overlap
appreciably with each other, the fusion process is governed by a single 
adiabatic one-body potential 
where the excitation 
on the adiabatic base is neglected. 
(3) The transition from the two-body
treatment to the one-body one takes place 
at near the touching configuration,
where all physical quantities are smoothly joined.

To this end, we adopt Yukawa-plus-exponential (YPE) 
potential~\cite{kra79} as a basic
ion-ion potential $V_N^{(0)}$, because the diagonal
part of this potential satisfies the conditions 
(1)-(3) by choosing a suitable neck-formed shape for the one-body
system, as has been shown in our previous work~\cite{ich07-1}.
In addition, the saturation property of the nuclear matter is
phenomenologically taken into account in the YPE model.
It has also been shown that the YPE model is consistent with the
potential obtained with the energy density formalism with the Skyrme
SkM$^{*}$ interaction~\cite{vaz81,de02}.

\begin{figure}
\includegraphics[keepaspectratio,width=\linewidth]{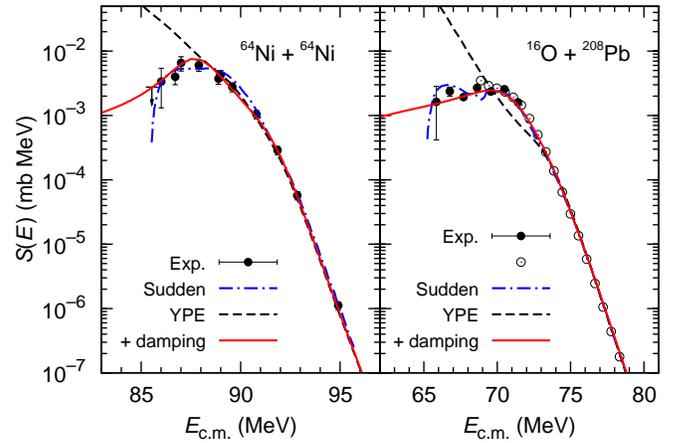}\\%
 \caption{\label{fig3}(Color online) Astrophysical S-factor
 for the $^{64}$Ni+$^{64}$Ni and $^{16}$O+$^{208}$Pb systems as a
 function of the incident energies. 
The meaning of each line is the same as in Fig.~1 
except for the dot-dashed line, 
which shows 
the result of
 the sudden model taken from Refs.~\cite{mis07} and \cite{Esb07}.
 }
\end{figure}

The nuclear coupling form factor which describes excitations to the
vibrational states in the two-body channel is taken as the
derivative of potential $V_N^{(0)}$~\cite{Esb87}. 
The coupling matrix elements are evaluated with the eigen-channel 
representation as in Eq. (24) in Ref.\cite{ccfull}. 
In order to satisfy the conditions
(1)-(3), we employ the following form for the nuclear potential
for the eigen-channel $\alpha$, 
\begin{eqnarray}
 V_N^{}(r,\lambda_\alpha)=V_N^{\rm (0)}(r)+\left[-\frac{dV_N^{\rm (0)}}{dr}\lambda_\alpha
+\frac{1}{2}\,\frac{d^2V_N^{\rm (0)}}{dr^2}\lambda_{\alpha}^{2}\right]\Phi(r,\lambda_\alpha), 
\end{eqnarray}
where $\lambda_\alpha$ is the 
eigen value of the excitation operator. 
The most important modification from the standard CC
treatment is the introduction of the damping factor $\Phi$. 
This damping factor represents the physical process for the gradual
transition to the adiabatic approximation, by diminishing the
strength of excitations to the target and/or projectile vibrational
states after the two nuclei overlap with each other.  
We thus choose the damping factor given by 
\begin{eqnarray}
\Phi(r,\lambda_\alpha)=
 \begin{cases}
  1 & \text{$r \geq R_d+\lambda_\alpha$~(Two-body region)},\\
  e^{-\frac{(r-R_d-\lambda_\alpha)^2}{2a_d^2}}&\text{$r<R_d+\lambda_\alpha$ (Overlap region)},
 \end{cases}           
\end{eqnarray}
where $R_d$ is the spherical touching distance between the target and
projectile defined by $R_d=r_d(A_{T}^{1/3}+A_{P}^{1/3})$, $r_d$ is the damping radius
parameter, and $a_d$ is the damping diffuseness parameter.
Notice that the touching point in the damping factor depends on 
$\lambda_\alpha$, 
that is, the
strength of the excitations starts to decrease at the different distance in
each eigen channel.

It is slightly complicated to take into account the effect of the damping 
factor on the Coulomb coupling. When different multipole components are 
present simultaneously, the eigen channels, which are 
introduced to evaluate the 
nuclear coupling matrix elements, do not diagonalize the Coulomb coupling 
matrix. We have therefore introduced the channel independent damping 
factor for the Coulomb coupling, but the effect on the 
fusion cross sections appeared small. 
For simplicitly, we therefore consider the damping factor 
only for the nuclear coupling in the calculations presented below. 

\begin{figure}
\includegraphics[keepaspectratio,width=\linewidth]{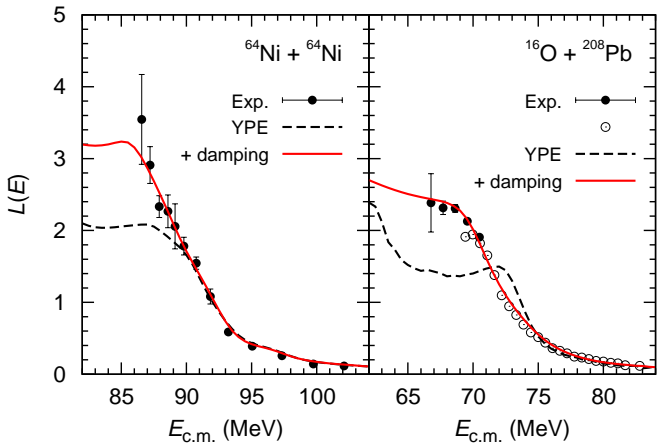}\\%
\caption{\label{fig4}(Color online) Logarithmic derivatives of
 fusion cross sections, $L(E)=
d\ln(E_{\rm c.m.}\sigma_{\rm fus})/dE_{\rm c.m.}$, 
for the $^{64}$Ni+$^{64}$Ni and
 $^{16}$O+$^{208}$Pb systems as a function of the incident energies. 
The meaning of each line is the same as in Fig.~1.
}
\end{figure}

We apply our present model to the fusion reactions for the
 $^{64}$Ni+$^{64}$Ni and $^{16}$O+$^{208}$Pb  systems.
To this end, we incorporate the damping factor and the YPE potential in 
the computer code {\tt ccfull}~\cite{ccfull}.
For the $^{64}$Ni+$^{64}$Ni system, the coupling scheme included in the 
calculation, as well as the deformation parameters, are the same as in 
Ref.~\cite{jiang04-2}.  
For the $^{16}$O+$^{208}$Pb system, those are the same as in
Ref.~\cite{Mor99}, but we include the coupling to the low-lying 3$^{-}$ phonon
states and the double-octupole phonon excitations for both the $^{16}$O
and $^{208}$Pb nuclei.
For the damping factor, we use $r_d=1.298$ fm and $a_d=1.05$ fm for the
$^{64}$Ni+$^{64}$Ni system, and $r_d=1.280$ fm and $a_d=1.28$ fm for the
$^{16}$O+$^{208}$Pb system. 

\begin{figure}[t]
\includegraphics[keepaspectratio,width=\linewidth]{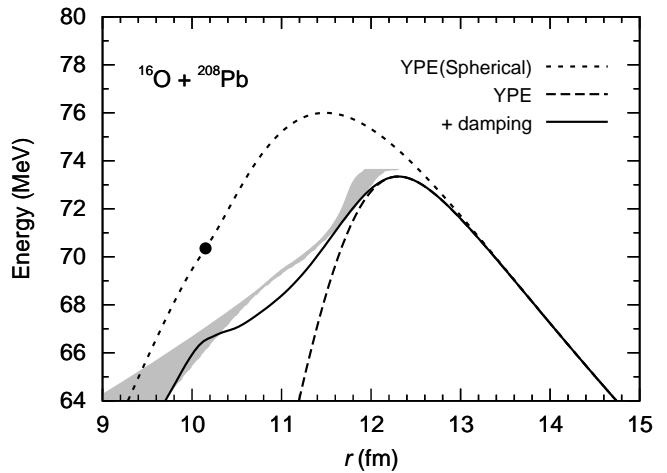}\\%
\caption{\label{fig5} The adiabatic potential for the $^{16}$O+$^{208}$Pb
 system as a function of the center-of-mass distance.
The solid line is the adiabatic potential obtained with the damping
 factor. The dashed line is the result obtained with the conventional
 CC approach. The dotted line and the solid circle are the pontential and the 
touching point for the uncoupled case, respectively. 
 The gray region denotes the adiabatic potential obtained with the
 potential inversion method, taken from Ref.~\cite{Hagi07}.
}
\end{figure}

For the YPE model,
the parameters are taken as ${a_0}=0.68$ fm, $a_{s}=21.33$ MeV, and
${\kappa_s}=2.378$ from FRLDM2002~\cite{mo04}. In order to fit the
experimental fusion cross 
sections, the radius parameter $r_0$ is adjusted to be 1.205 fm
and 1.202 fm for the $^{64}$Ni+$^{64}$Ni and $^{16}$O+$^{208}$Pb
systems, respectively.
For the mass asymmetric $^{16}$O+$^{208}$Pb system,
it is difficult to joint smoothly the potential energies between the
two-body and the adiabatic one-body systems at the touching point,
because the proton-to-neutron ratio for the one-body system differs from
that for the target and projectile in the two-body system. 
To avoid this difficulty, we smoothly connect the potential energy
around the touching point to the liquid-drop energy of the compound
nucleus, using the third-order polynomial function (see.~the dashed line
in Fig.~4).  
We do this by identifying the internucleus 
distance $r$ with the centers-of-masses distance of two half spheres.  
The obtained potential is similar to the result of the density-constrained
time-dependent Hartree-Fock method~\cite{Umar07}. 
We have checked this prescription for the mass symmetric
$^{64}$Ni+$^{64}$Ni system, 
by comparing to the potential energy
used in our previous work~\cite{ich07-1}. 
The deviation due to this prescription is 
negligibly small.

Figure 1 shows the fusion cross sections thus obtained.
The fusion cross sections obtained with the
damping factor are in good agreement with the
experimental data for both the systems (see the solid line).
The dashed line is the YPE potentials without the damping factor.
The dot-dashed line is the results of no coupling with the YPE potential.
For both the systems, we see that drastic improvement has been achieved
by taking into account the damping of the CC form factors.

We also compare the astrophysical $S$ factor representation of the
experimental data with the calculated results, as shown in Fig.~2.
In the calculation, the Sommerfeld parameter $\eta$ 
is shifted by 75.23 and 49.0 for the $^{64}$Ni+$^{64}$Ni 
and $^{16}$O+$^{208}$Pb systems, respectively.
The $S$ factor obtained with the damping factor are consistent with the
experimental data for both the systems (see the solid lines), and
reproduce well the peak structure.

For astrophysical interests, it is important to evaluate fusion cross
sections at extremely low incident energies, which is difficult to
measure directly.
Notice that the $S$ factor predicted by our model 
differs considerably from that of the sudden
model by Mi\c{s}cu and Esbensen~\cite{mis07}, denoted by the
dot-dashed line.
For both the systems, as the incident energy decreases, their $S$ factor
falls off steeply below the peak of the $S$ factor, while our $S$ factor
has a much weaker energy dependence.  

Figure 3 compares the logarithmic derivatives 
$d\ln(E_{\rm c.m.}\sigma_{\rm fus})/dE_{\rm c.m.}$ of the experimental
fusion cross section with the calculated results.
It is again remarkable that only the result with the damping factor 
achieves nice reproduction of the experimental data. 
For the $^{64}$Ni+$^{64}$Ni system, the result with the damping factor
becomes saturated below $E_{\rm c.m.}$=86 MeV. 
This behavior is similar to the experimental data for 
the $^{16}$O+$^{208}$Pb system. 
The measurement at further lower incident
energies for this system 
will thus provide a strong test for the present adiabatic model. 

Figure 4 shows the adiabatic potential of the $^{16}$O+$^{208}$Pb
system, that is, the lowest eigenvalue  
obtained by diagonalizing the coupling matrix at each
center-of-mass distance $r$. 
The adiabatic potential calculated with and without the damping factor are
denoted by the solid and the dashed lines, respectively.
The uncoupled YPE potential 
is also shown by the dotted line. The solid circle
denotes the touching point in the absence of channel coupling.
We see that the result obtained with the damping factor is much thicker
than that of the conventional CC model.
In this respect, it is interesting that the result with the
damping factor is 
similar to that obtained with the potential inversion
method~\cite{Hagi07}, denoted by the gray region, 
justifying our treatment for the damping of the CC form factor.

For the average angular momentum of the compound nucleus, the results
with the damping factor become saturated at incident energies below the
threshold energy with decreasing incident energy, as shown in our
previous works~\cite{ich07-1,ich08}.
This result largely differs from that obtained with the sudden model by
Mi\c{s}cu and Esbensen. Their result is 
strongly suppressed at energies below the threshold energy. 
It is thus interesting to measure the average angular momentum at deep 
subbarrier energies, in order
to discriminate the two approaches.

We have also applied our model to the $^{58}$Ni+$^{58}$Ni reactions and
the results obtained are in good agreement with the
experimental data. 
For the damping factor, we used $r_d=1.3$ fm and
$a_d=1.3$ fm in order to fit the experimental data.
Notice that 
the obtained damping radius parameters for the three
systems which we study are almost the same.
We emphasize that our model achieves an overall fit not only to the
fusion cross sections but also to the S factors and the logarithmic
derivatives simultaneously.

In summary, we have proposed 
a novel coupled-channels approach for heavy-ion fusion reactions 
by introducing 
the damping of the CC form factor 
inside the touch point 
in order
to simulate the transition from the diabatic to adiabatic states.  
The important point in our present model is that the transition
takes place 
at different places for each eigen channel. 
By applying this model to the $^{16}$O+$^{208}$Pb, the $^{64}$Ni+$^{64}$Ni, and
 the $^{58}$Ni+$^{58}$Ni systems, we conclude that the smooth transition from
the diabatic two-body to the 
adiabatic one-body potential is responsible for the
steep falloff of the fusion cross section. 
It is an interesting future study to apply the present model 
systematically to other systems and clarify the dynamics of 
deep subbarrier fusion reactions, and thus many-particle tunneling 
phenomena. 

\begin{acknowledgments}
 T.I. thanks H. Feldmeier for useful discussions.
This work was supported by the Japanese
Ministry of Education, Culture, Sports, Science and Technology
by Grant-in-Aid for Scientific Research under
the program number 19740115.
\end{acknowledgments}

\end{document}